\documentclass[12pt,a4paper,twoside]{paper}
\usepackage{amsmath,bbm,amssymb,amsxtra}
\usepackage{hyperref}
\usepackage{graphicx}
\usepackage{epstopdf}
\usepackage{chngcntr}
\usepackage{etoolbox} 
\usepackage{times}
\def\R{\mathbb R}

\newcommand{\NX}{{{\bf X}}}

\newcommand{\Nx}{{{\bf x}}}

\begin{document}

{\Huge Mind, Matter, and Freedom in Quantum Mechanics and the de Broglie-Bohm Theory} \vspace*{5mm}

{\LARGE Valia Allori
\vspace*{5mm}
 
 Universit{\'a} degli Studi di Bergamo, via Pignolo 123, 24121 Bergamo, Italy; }\footnote{E-mail:valia.allori@unibg.it}
\vspace*{5mm}

{\LARGE Jean Bricmont

\vspace*{5mm}
 
 IRMP, UCLouvain, chemin du Cyclotron 2, 
1348 Louvain-la-Neuve
Belgium}\footnote{E-mail: jean.bricmont@uclouvain.be}
 \vspace*{5mm}


\begin{abstract}
There are several important philosophical problems to which quantum mechanics is often said to have made significant contributions:
\begin{itemize}
\item Determinism: quantum theory has been taken to refute determinism;
\item Free Will: in turn, this is thought to open the door to free will;
\item The mind-body problem: relatedly, it is sometimes said to shed light on consciousness;
\item Idealism: more radically, quantum theory is assumed to have refuted realism and to have placed the observer at the center of the world;
\item Reductionism: even granting realism, it has been claimed that quantum theory undermines reductionism.
\end{itemize} 
Our main thesis in this paper is that none of this is either necessary or desirable. By adopting the de Broglie–Bohm theory (or Bohmian mechanics), one can straightforwardly account for quantum phenomena without endorsing any of these claims.
\end{abstract}
\begin{keywords}
quantum theory; de Broglie-Bohm theory; realism; idealism; reductionism; determinism; predictability; free will; the mind-body problem; consciousness
\end{keywords}
\tableofcontents

\section{ The Orthodox and de Broglie--Bohm Formulations of Quantum Mechanics}\label{sec1}
Even if people often say ``quantum mechanics,” there is no unique theory able to account for quantum phenomena. In this paper we will focus only on two quantum theories: the one found in physics textbooks, usually called ordinary, or standard, or even orthodox quantum mechanics, and the one developed by de Broglie and Bohm, therefore dubbed the de Broglie—Bohm theory. 

Both orthodox quantum mechanics and the de Broglie--Bohm theory reproduce the same empirical phenomena and share much of the same mathematical formalism, including the wave function and the use of operators. Despite these formal similarities, they differ radically in their conceptual implications, as we will see.

\subsection{Ordinary Quantum Mechanics }\label{sec1.1}

Orthodox quantum theory emerged somewhat chaotically during the first two decades of the 20$^\text{th}$ century, developed by physicists such as Heisenberg, Born, Jordan, Bohr, Schr\"odinger, Pauli, and Dirac. This period of rapid innovation culminated in an axiomatic formulation in 1932 by von Neumann, providing a more systematic foundation for the theory \cite{vN}.
In this formulation, the complete physical description of any system at a given time (its instantaneous ``state'') is given by its wave function $\Psi$, introduced by Schr\"odinger.\footnote{Strictly speaking, one should say its quantum state, which may also include spin states, but we will ignore this distinction here.} 
The wave function evolves in time in two different ways: 
\begin{itemize}
\item In the absence of measurements, it obeys Schr\"odinger's equation, which is deterministic and continuous in time;
\item Whenever a measurement is performed on the system, $\Psi$ follows the so-called collapse rule: it jumps randomly and discontinuously, and the measurement result is determined by that jump, as described below.
\end{itemize}	
It is thanks to this second law of evolution that the wave function acquires a physical meaning in ordinary quantum theory. 
In fact, ordinary quantum theory assumes that physical quantities such as position, momentum, and energy, which can \textit{a priori} take measurable values, are represented by mathematical objects called self-adjoint operators. Experimental results are then given by the operator’s eigenvalues corresponding to the eigenvector into which the wave function has ``jumped.''\footnote{
An eigenvalue equation has the form $\hat{A}x = \lambda x$, where $\hat{A}$ is an operator (or a matrix), $x$ a vector, and $\lambda$ a number called respectively the eigenvector and the eigenvalue. This means that applying the operator to the vector $x$  yields the same vector rescaled by that number. In quantum theory, operators act on states, and the eigenvalue equation links an operator, a wave function, and the measurable value of an observable. Written as $\hat{A}\Psi = \lambda\Psi$, it says that $\Psi$ is an eigenstate of the observable $\hat{A}$ with eigenvalue $\lambda$, namely the possible measurement outcome. The most famous example is the time-independent Schr\"odinger equation $\hat{H}\Psi=E\Psi$, where the Hamiltonian operator $\hat{H}$, representing energy, determines the allowed energy levels of a quantum system.}
There is an algorithm called \textit{the Born rule}, which allows us to compute the probabilities of the values observed as a result of a measurement of a physical quantity represented by the operator $\hat{A}$  when the system's state is $\Psi$. More specifically, the Born rule states that the probability of obtaining a given outcome $\lambda$ is the square of the absolute value of the coefficient of the corresponding eigenvector in the expansion of the wave function in the basis of eigenvectors of $\hat{A}$ (which exists quite generally).

Despite its remarkable empirical success, this formulation of quantum mechanics has, from the outset, raised several natural questions:
\begin{itemize}
\item Why do we have two laws of evolution and not one? Why are measurements treated differently from other physical processes?
\item Relatedly, why is the notion of ``measurement'' so central to the very definition of a physical theory? 
The same applies to the notion of ``observer,'' since one can reformulate the second law of evolution as: ``Whenever an observer interacts with the system, $\Psi$ jumps [\dots].''
\item And finally, what happens outside of modern laboratories, or before they existed, or even before humans appeared and began making measurements in the first place? 
\item Because of that, ordinary quantum mechanics should be viewed, strictly speaking, as a theory that predicts what happens in some laboratory operations and nothing else. This is, to say the least, puzzling, as John Bell remarked: ``But experiment is a tool. The aim remains: to understand the world. To restrict quantum mechanics to be exclusively about piddling laboratory operations is to betray the great enterprise. A serious formulation [of quantum mechanics] will not exclude the big world outside the laboratory."  \cite[p. 34]{Be2}
\end{itemize}

Moreover, notice that there is no mention about what matter is: the rules above merely tell us how to extract empirical predictions from the formalism. 
Nonetheless, orthodox quantum theorists often use the word ``particle,” even if they also warn us not to take the terminology too literally. In fact, they have a definite position and a definite velocity only when they are observed and, given 
Heisenberg’s uncertainty principle, particles cannot have a definite value of position and momentum at the same time. Thus, they cannot have trajectories. 

Such difficulties are among those which have motivated the search for alternative theories, chief among them the de Broglie--Bohm theory, to which we now turn.
	
\subsection{The de Broglie-Bohm Theory }\label{sec1.2}
Between 1923 and 1927, de Broglie proposed a quantum theory in which measurements and observations play no fundamental role \cite{dB10, BV}. Due to historical circumstances, he later abandoned it (see e.g. \cite{BV}), but in 1952 Bohm independently reproposed a version of de Broglie’s theory \cite{Bo1}. It has since become known as the de Broglie-Bohm theory.\footnote{
The theory goes by several names, such as the pilot-wave theory or Bohmian mechanics.
}
Bell \cite{B} contributed significantly to its popularization, along with many others, see e.g. \cite{Al, alloribook, Bri6, DGZ, DL, DT, Ho, Ma10, Norsen, Tow, Tu3}.

Here is a brief sketch of the theory. In contrast to ordinary quantum theory, matter in the de Broglie-Bohm theory is composed of point-like particles with definite positions at all times, and therefore well-defined trajectories, independent of observation. The complete description of a system includes both the wave function and the particles’ positions. For a system of $N$ particles at time $t$, the state is specified by $\big(\Psi (t), {\bf X}(t)\big)$, where $\Psi (t)=\Psi(\Nx_1, \dots, \Nx_N, t)$ and ${{\bf X}(t)}= \big(\NX_1(t),\ldots, \NX_N (t)\big)\in\R^{3N}$ are the actual positions of the particles.
The evolution of the state $(\Psi , \bf X)$ is governed by two deterministic laws:
\begin{itemize}
\item $\Psi$ obeys the Schr\"odinger equation at all times: the wave function never collapses, though it effectively appears to do so in suitable circumstances;
\item Particle positions evolve according to their velocities, which depend on the wave function evaluated at the positions $\big(\NX_1(t),\ldots, \NX_N (t)\big)$ of all particles at time $t$.  
\end{itemize}

Even though the evolution is deterministic, the initial particle positions may be unknown. As in classical statistical mechanics, one can assume a probability distribution. Instead of a uniform distribution, the ``quantum equilibrium'' distribution $\rho=|\Psi|^2$ is used due to its property of {\it equivariance} \cite{DGZ}, which means that, if particle positions are initially distributed by $|\Psi|^2$, they continue to do so at all later times. Under this assumption, the de Broglie-Bohm theory reproduces all empirical predictions of ordinary quantum mechanics.

While both orthodox quantum mechanics and the de Broglie-Bohm theory reproduce the same laboratory predictions, they differ sharply in their theoretical status. Indeed, de Broglie-Bohm provides a clear account of particle positions and trajectories, eliminating the special role of measurement that is central to the orthodox formulation.

In this theory ``measurements'' are simply physical interactions between a quantum system and an apparatus. Importantly, except for position measurements, they generally do not reveal pre-existing properties of the system,\footnote{There are theorems, due to Bell \cite{Bell} and to Kochen and Specker \cite{KS} showing that one cannot assume, in general, that ``measurements" reveal pre-existing properties of the system being ``measured".} because they are genuine interactions between the system and the measuring device, which means that the result of that ``measurement" (quite a misnoner in this case, see \cite{Be2}) does not depend only on the properties of the system, but also  on the detailed properties of the measuring apparatus. 

\section{Determinism, Free Will and Consciousness}\label{sec3}
Determinism is the view that, given the state of the world at a particular time, both its future and past states are fixed. If humans are part of this world, and the world is deterministic, then our actions are determined just like everything else. If that is the case, nothing we do is truly ``up to us,” and we cannot be considered free. Quantum theory has sometimes been invoked to challenge this view, since its laws are indeterministic. 
Yet, some have rightly pointed out that free actions cannot simply follow laws—deterministic or stochastic—they require self-awareness and intentionality, capacities provided by consciousness. Quantum theory has also been interpreted by many as granting consciousness a crucial role in this regard. In this section, we will explore these issues in more detail.

\subsection{Determinism vs. Predictability} 
In classical mechanics, the state of a system is given by its instantaneous position and velocity. The theory is deterministic: given the complete state of a system at one time, laws specify its state at all times.
This implies that the temporal evolution of every single physical body is fully determined since the beginning of the universe. 
This applies to all physical systems: a tossed coin, a thrown die, or a roulette ball. 

It is important to notice that determinism does not imply practical predictability, even though the concepts are often confused. Determinism implies \textit{in principle} predictability: given the exact initial state, one could determine how a system will evolve.
 Laplace illustrated this with an ``intelligence'' possessing all information and computational power to solve the equations of motion: ``[\ldots] for it, nothing would be uncertain and the future, as the past, would be present to its eyes'' \cite[p.~4]{La}. However, this does not mean prediction is possible in practice, since we lack complete information and computational capacity.\footnote{Laplace added that {\em we\/} shall ``always remain infinitely removed" from this imaginary ``intelligence.'' \cite[p.~4]{La}.} We might not know the initial conditions, but we also might not even know the laws: for example, the planets obeyed the laws of gravitation long before we knew them.  

Determinism concerns the nature of the (possibly unknown) laws governing a phenomenon, while practical predictability depends on our ability to know and apply these laws. 
Our limitations are what make probabilistic predictions necessary in practice. For example, the uncertainty in the result of tossing a coin   arises not from the laws themselves but from our imperfect knowledge. In this sense, probability within a deterministic framework is epistemic rather than ontic.

One might think that, if the laws are known, then a perfect knowledge of initial conditions would be unnecessary for approximate predictions. Yet sensitive dependence on initial conditions, or chaos, shows this is not always the case. Small differences in initial states can grow exponentially, making long-term predictions practically impossible. Weather provides a classic example, summarized in the ``butterfly effect:'' the flap of a butterfly’s wings in Brazil might trigger a tornado in Texas. Similarly, in billiards, tiny variations in velocity or point of impact can drastically alter outcomes, such as whether the ball falls into a pocket or strikes ball number 8. 

These examples underline the distinction between determinism and practical predictability: systems can be deterministic yet unpredictable in practice. Thus, unpredictability does not imply indeterminism.\footnote{For example, in an often quoted lecture
to the Royal Society, on the three hundredth anniversary of Newton's {\it Principia\/}, the distinguished British mathematician Sir James Lighthill
 gave a perfect example, based on unstable dynamical systems,  of how to confuse  predictability and determinism: ``we are all deeply conscious today that the enthusiasm of our forebears for the marvelous achievements of Newtonian mechanics led them to make generalizations in this area of \textit{predictability} which [\dots]  we now recognize were false [\dots]  about \textit{determinism} of systems satisfying Newton's laws of motion that, after 1960, were to be proved incorrect" \cite{Li}. ``After 1960'' refers to the study of unstable or chaotic deterministic systems. }  Other arguments are required to support indeterminism, which is where quantum mechanics has been invoked.

\subsection{Determinism and Quantum Theories}\label{sec3.2}
Newton’s celestial mechanics, describing the motion of massive bodies, is the archetypical example of deterministic laws. All pre-quantum theories are likewise deterministic. In the 19th century, Maxwell’s classical electrodynamics showed that charged particles generate electromagnetic waves, guiding their motion through deterministic interactions. In the early 20th century, the special and general theories of relativity, developed by Lorentz, Poincar\'e, Einstein, and Hilbert, revised Newton’s laws of motion and gravitation but retained determinism.\footnote{Several caveats are necessary for full rigor, but they lie beyond the scope of this article.}
The real break with determinism in physics came with quantum mechanics. The formalism of ordinary quantum mechanics incorporates indeterminism at a fundamental level: one associates states to physical systems, and given a state, one can only compute the probabilities of transitioning to another state upon a ``measurement.''  

In contrast to classical physics, where probabilities are epistemic, in ordinary quantum theory they are ontic: there is no information that would allow us to predict which state will result, even in principle. Not even a Laplacian Demon could determine the outcome of a measurement. Thus, quantum mechanics provides the first fundamental physical theory governed by intrinsically indeterministic laws.

Now, let us contrast this with the de Broglie—Bohm theory. In this context, the indeterminism of the quantum predictions is not fundamental. The use of probabilities is not due to a lack of deterministic laws, but rather about the impossibility of predicting determinate outcomes. Thus, probabilities are  epistemic. 
As we saw, we use epistemic probabilities in classical physics when we do not know precisely the system’s initial state. In the de Broglie-Bohm theory the situation is similar but in a sense worse, because this ignorance is ineliminable:  we never know the exact location of the particles. This uncertainty is due to quantum equilibrium. To come to gain information about an unknown system we need to interact with it, and to do this we need not to be in equilibrium with it.\footnote{For instance, a thermometer measures body temperature by interacting with the system until equilibrium is reached. It is designed to minimally disturb the system while its own state changes to display information. Its thermal capacity and calibrated scale ensure the reading closely reflects the actual temperature. A ``perfect thermometer''—showing the exact temperature without interaction—is impossible, but the closer the reading is to the true value, the more precise it is, and in most cases, the interaction can be safely neglected.} 

But if, in the de Broglie-Bohm theory,  everything is in quantum equilibrium, there is nothing out of equilibrium that could be used to measure particle positions with more precision than what we get from the $|\Psi|^2$ distribution. This ``absolute uncertainty'' underlies probabilistic predictions: it is impossible to obtain enough information for the deterministic equations to yield precise particle locations. Consequently, all quantum systems are deterministically unpredictable—one can only specify probabilities of outcomes, since deterministic prediction would require knowledge of the exact initial conditions.
Notice that this does not mean that particle locations are not knowable in principle: an all-knowing Laplacian demon could still have access to them, because by definition he would not need to interact with anything to get this knowledge.  Nonetheless, given that we do need to interact to get information, there are serious limits about what we can predict.  Hence, even if the  setting is deterministic, the experimental outcome are described probabilistically.

\subsection{Determinism and Free Will}\label{sec3.3}
Be that as it may, the notion of universal determinism of physical laws has provoked much hostility because it seems to contradict our notion of free will. Thus, some have rejoiced in the idea of an indeterministic quantum world, as it would seem to leave space for freedom. 
To discuss this, one must first try to define free will. It is not just the feeling that I may sometimes act without external constraints, without anybody forcing me to do or not do something, but that  ``I" choose to do one thing rather than another. When I am presented a choice, what I end up doing is ``up to me."
The choice can be trivial, like choosing between having ice cream or cake for dessert, or serious, like choosing which profession to embrace or whether or not to commit a crime. But in our everyday life, we constantly feel that it is possible to make at least some choices under no external constraint. In that case, we choose freely; we act on our  own free will.

In summary, the idea of free will in tension with determinism is connected with the idea of control: if we control the outcome of an action, then we act freely; otherwise, if the action is beyond our control, the action is not free. Consequently, if determinism is true, there cannot be free will. 

Schematically, the argument from determinism against free will has three premises \cite{vIW}: 
\begin{itemize}
\item P1: determinism is true; 
\item P2: we act freely only if we have control over our actions; 
\item P3: if  determinism is true, then we have no control over our actions; 
\end{itemize}
This leads to the conclusion that \begin{itemize}
\item 4 (from P1 and P3): we have no control over our actions, 
\item 5 (from P2 and 4): we are not free. 
\end{itemize}
This conclusion is contested, as we ``obviously” have free will! 

To make perhaps things worse, free choices are closely linked to moral responsibility, which has also legal implications. Courts distinguish between those responsible for a crime and those who are not: someone who murders out of revenge is held accountable, whereas a person committing murder due to hallucinations is not. Responsibility is typically tied to actions arising from free will.  However, if determinism is true, our desires and intentions are also determined by physical laws. Hard determinism\footnote{Which is a philosopher's notion, meaning determinism applied to our actions, wishes, intentions etc.} is the view that no one is ever truly free in this sense: we have no control over our actions. Thus, there is no fundamental difference between criminals and those acting without apparent choice. Both are like computers executing programs, with no control over the laws governing their neurons.\footnote{Here we discuss determinism, not practical predictability.} Consequently, in this view, no one is genuinely responsible for their actions, and the usual admiration or blame for moral deeds is, strictly speaking, misplaced (although they are also determined by physical laws).

\subsection{Freedom from the Quantum?}
It has been argued that ordinary quantum mechanics offers a way to avoid these unwelcome conclusions—namely, that we have no genuine free will, and consequently no justification for blaming criminals or admiring heroes.\footnote{For example, Gisin has discussed the link between the lack of determinism in quantum mechanics and free will, see \cite{Gis1, Lyle}.} 
 Since the theory posits truly indeterministic laws, the argument from universal determinism to ``free will is an illusion'' fails, as its first premise is false. In other words, if not everything is determined, there is at least the possibility for genuine choice. This also opens the door to grounding moral responsibility in physical reality: actions may not be fully predetermined, so individuals can genuinely influence outcomes, justifying praise, blame, punishment, or reward. 

Nonetheless, many have noted that showing that the above argument fails does not establish that we can actually behave freely. 
Hard determinists have maintained that, if genuine free will requires control, as emphasized by P2, then we have no control over any laws, whether deterministic or random. In a deterministic universe, I am determined to type these words; it is impossible for me to do otherwise, even if I feel I could \cite{pere}. In an indeterministic world I might end up typing something different, for instance with a $40\%$ chance of the sentence I actually write and $60\%$ for an alternative. The future is genuinely open.  Yet this does not help with free will. In fact, although outcomes are not certain, which possibility occurs is still not under my control—the laws govern the result, not me. Hence, there is no room for a free action neither in a deterministic nor in an indeterministic universe. 

Compatibilists instead argue for another notion of free will for which determinism is required. A free action is not an uncaused action as hard determinist have suggested. Rather, an action requires determinism to be labelled as free: it needs to be caused, but it also needs to  be caused in the ``right” way. That is, a free action is an action which is not forced upon us but rather it is caused by our desires, beliefs and intentions \cite{dennett}. If so, indeterminism does not help at all with freedom. 

In contrast with compatibilists, libertarians about free will (not to be confused about libertarians about economics) agree with hard determinists that free will is about control and thus is incompatible with determinism. Nonetheless, as already stated, we have no control over 
laws, whether they are deterministic or not. 
This however, is the case if one assumes that we are simply complicated machines obeying laws, and some libertarians deny this. For this reason the problem of free will is closely tied to what philosophers call ``the mind-body problem,' which complicates the matter. Let us see how in the next section. 

\subsection{ The ``Mind-Body Problem"}\label{sec4}
Libertarians argue that some actions—those we call free— are not determined but they are genuinely ''coming" from  us through our agency \cite{kane}. This is sometimes seen as grounded in consciousness understood as irreducible to the material body \cite{descartes, swinburn}.\footnote{Descartes wrote:`` “I experience in myself a certain freedom of the will, by virtue of which I can refrain from or pursue what the intellect proposes, in such a way that I am not constrained by any force.”}

This is connected to one of the central debates in philosophy of mind is the relation between mind and body. Some argue that the body made of matter, obeys laws (deterministic or not), while the mind does not. We experience thoughts, feelings, and choices, which appear fundamentally different from the rest of the world, which seems mechanistic. For some, the best explanation for this is that they actually are fundamentally different: there is a mind which allows us to have conscious experiences. 
Nagel’s famous argument in ``What is it like to be a bat?" \cite{Nagel1} highlights a problem of reducing the mind to the body: 
while I can guess what it feels like for another person to see red by analogy with my own experience, no such analogy is possible with a bat’s echolocation. No matter how much we know about the bat’s brain or body, we gain no insight into what it \textit{feels like} to be a bat. This reveals a gap between the subjective, qualitative character of experience and the objective, quantitative accounts of physical processes. That gap is one aspect of the ``mind-body problem," also called the ``hard problem of consciousness" \cite{chalmers}, or the problem of \textit{qualia}.

In what follows, we set aside the question of free will and turn instead to the broader issue of how quantum theory has been invoked to shed light on the nature of consciousness.

\subsection{Consciousness and the Quantum}
Some have suggested that quantum theory may provide a framework for understanding features of consciousness that resist reduction to classical physical processes. One possible link is the claim that the collapse of the wave function is produced by a non-physical consciousness, as suggested quite explicitly by Eugene Wigner (1902--1995), co-recipient of the Physics Nobel Prize in 1963.\footnote{He wrote: ``[\dots] It will remain remarkable, in whatever way our future concepts may develop, that the very study of the external world led to the scientific conclusion that the content of the consciousness is an ultimate reality. [\dots] The preceding argument for the difference in the roles of inanimate observation tools and observers with a consciousness --- hence for a violation of physical laws where consciousness plays a role --- is entirely cogent so long as one accepts the tenets of orthodox quantum mechanics in all their consequences.” \cite{Wi1}.} He was convinced that it was ``not possible to formulate the laws of quantum mechanics in a fully consistent way without reference to the consciousness'' \cite[p. 169]{Wi1}.\footnote{Nonetheless, Wigner’s view on the role of consciousness in quantum mechanics evolved over time (see \cite{Esf}).}  

As we have already seen, however, this is not the case: it is indeed possible to reproduce quantum phenomena without invoking consciousness, as evident in the de Broglie–Bohm theory, where the wave function always evolves deterministically. Under circumstances in which, in ordinary quantum theory, the wave function collapses, in the de Broglie–Bohm theory one can treat the wave function \textit{as if} it had collapsed, in the sense that the relevant information for predicting the system’s evolution is contained in one term of the superposition, and the others can be effectively ignored.  

A completely different connection between consciousness and quantum mechanics has been proposed by Roger Penrose (Nobel Prize in Physics in 2020) \cite{Penr, Penr1}. He argues that the human mind exhibits certain non-computational aspects—namely, it is capable of reasonings that no machine, however sophisticated, could possibly reproduce—and that this feature can be explained quantum mechanically.\footnote{See \cite{BDU} for a discussion of this and related ideas from a physics perspective.} Penrose’s arguments are highly technical, controversial, and speculative. Even if they were correct, however, they would not resolve the problem of consciousness. For that would require bridging the gap between our subjective sensations and our objective view of the world, whereas quantum mechanics lies entirely on the ``objective'' side of the divide.  

To conclude, therefore, as far as we can see, quantum mechanics is unlikely to resolve the riddle posed by the relation between the body and the conscious mind. Let us now turn to the more relevant (for quantum mechanics) topic of realism, and its connection with explanation and reductionism.

\section{ Realism and Reductionism }\label{sec2}
Many practicing scientists who have not studied quantum mechanics would likely say that scientific theories are approximately true: they provide a fallible but generally reliable picture of the world. This view is known as scientific realism, which presupposes realism, namely the thesis that there is a world independent of our minds. After the advent of quantum theory, however, some physicists have argued that scientific realism is impossible and that even the more basic notion of realism is in jeopardy.  

In this section, we will examine the reasons behind these radical claims and show that they are largely unwarranted. We will also discuss alternative positions suggesting that quantum theory can be made compatible with realism. Assuming realism to be true, however, there has been disagreement about what the theory reveals about the world. Some have even argued that one must abandon reductionism—the idea that macroscopic phenomena can be fully explained in terms of fundamental microscopic entities and their dynamics. To conclude, we will show that this is neither true nor desirable. 

\subsection{Realism  vs. Idealism }
Realism is the view that an objective reality exists independently of our minds, and that it is knowable through experience. This commonsensical perspective underlies everyday life and scientific practice: if we see chairs or tables, we naturally assume they exist. Realism assumes that the senses are generally reliable.  

Idealism, by contrast, asserts that reality is fundamentally mental: our knowledge of the world comes from within ourselves rather than from the world ``out there.'' Its skeptical critique points out that, since we access the world only through our perceptions, we cannot be certain that these perceptions correspond to external objects. Solipsism, the most extreme form, claims that nothing exists outside one’s own mind. The Irish bishop George Berkeley (1685--1753) summarized this view: ``to be is to be perceived.''  

While no one can decisively disprove idealism or solipsism, there are also no compelling positive arguments for them. As the Australian philosopher David Stove  (1927--1994) notes, idealists often rely on what he calls the \textit{gem} of idealism:\footnote{``You cannot have trees-without-the-mind in mind, without having them in mind. Therefore, you cannot have trees independent of the mind in mind" \cite[p. 139]{Stove}.} the claim is that because our perception of a tree requires our mind, the tree cannot exist independently. However, this reasoning is a \textit{non sequitur}. Berkeley also used ``the gem,” \footnote{He wrote: ``The mind [\dots] is deluded to think it can and does conceive of bodies existing unthought of, or without the mind, though at the same time they are apprehended by, or exist in, itself" \cite[p. 270]{BB}.} but also some prominent scientists like the French mathematician Henri Poincar\'e (1854--1912)  adhered to it.\footnote{See the following passage: ``All that is not thought is pure nothingness; since we can think only thought and all the words we use to speak of things can express only thoughts, to say that there is something other than thought, is therefore an affirmation which can have no meaning“ \cite[p. 355]{HP}.  }

Idealism is inconsistent, radical, and implausible. It fails to answer basic questions: who or what is the perceiver—the individual, humanity, God, or animals? Solipsism, meanwhile, is extreme and conspiratorial. Euler famously dismissed it with the ``incredulous stare’’ argument: would one truly deny the existence of a real world outside one’s mind?\footnote{“Thus when my brain excites in my soul the sensation of a tree or of a house I pronounce without hesitation that a tree or a house really exists out of me of which I know the place, the size and other properties. Accordingly we find neither man nor beast who calls this truth in question. If a peasant should take it into his head to conceive such a doubt, and should say, for example, he does not believe that his bailiff exists, though he stands in his presence, he would be taken for a madman and with good reason; but when a philosopher advances such sentiments, he expects we should admire his knowledge and sagacity, which infinitely surpass the apprehensions of the vulgar” \cite[pp. 428--449]{Eul}.}  

The most compelling argument against idealism seems to us the one put forward first by Locke: when seeking to account for our perceptions, we should favor the view that best explains them. Realism succeeds here: the coherence of our senses, such as seeing and touching a flat surface, is explained by an actual object producing both sensations. Idealism cannot account for this systematic coherence. Accordingly, we are justified in believing in realism rather than idealism.\footnote{ Locke wrote: ``The ideas of sense are not fictions of our fancies, but the effects of things operating on us without us, and really existing. For the testimony of our senses, joined with the constant and regular coherence of our perceptions, carries with it an assurance of the reality of external objects; since it is not in the power of imagination to frame such orderly and coherent appearances, which all the senses witness at once” \cite{locke}, Book IV, Chapter XI, 4.}

\subsection{Idealism and Quantum Mechanics}\label{sec2.2}
Nonetheless, quantum mechanics is often quoted as having given new life to idealism, because it is taken to be incompatible with realism. This belief has historical and sociological roots, which we summarize here.\footnote{For details, see \cite{alloriwhatif}, and references therein.}

Early quantum experiments and theoretical developments suggested that matter and radiation could not be adequately described by the classical categories of waves or particles. Under the influence of Bohr and the Copenhagen school, many were led to believe that realism had to be abandoned.
In 1926, Schrödinger attempted to treat matter as fundamentally a wave, introducing a wave function $\Psi$. Yet $\Psi$ is defined on the configuration space $\R^{3N}$ of $N$ particles, not on ordinary three-dimensional space, making it difficult to assign it a straightforward physical interpretation: $\Psi = \Psi(\Nx_1, \dots, \Nx_N, t)$, $\Nx_i \in \R^3$, $i=1,\dots,N$.\footnote{For $N=1$, $\Psi$ lives in $\R^3$, but this special case can be misleading.} Schrödinger himself rejected a literal wave ontology, as did others.\footnote{In fact, de Broglie called it ``paradoxical to construct a configuration space with the coordinates of points that do not exist” \cite{BV}, and Einstein found it ``does not smell like something real” \cite{howard}. In addition, Lorenz, while praising Schr\"odinger for the visualizability of his wave mechanics, wrote him that he would go back to Heisenberg’s matrix mechanics if one were to take the wave function as a physical field \cite{prizbram}.}
Around the same time, Heisenberg proposed the uncertainty principle, initially as a statement about epistemic limits, but interpreted ontologically, it implied that particles could not have precise positions and velocities simultaneously.\footnote{Heisenberg wrote:`` This uncertainly principle specifies the limits within which the particle picture may be applied. Any use of the words `position’ and `velocity’ with an accuracy exceeding that given by equation (the uncertainty principle) is just as meaningless as the use of words whose sense is not defined” \cite{heisPoP}, p. 15.}
Experiments further complicated the picture: electrons exhibited interference patterns, characteristic of waves, while the photoelectric effect supported a particle description of light. These results suggested that classical notions of particles and waves could not describe the quantum world.
Bohr’s principle of complementarity emerged to address this: certain properties, such as wave and particle aspects, cannot be observed simultaneously. Each manifests only under specific experimental conditions, and both perspectives are needed for a complete account of quantum phenomena.\footnote{In his Como lecture Bohr wrote: ``The wave picture and the particle picture, in their application to atomic phenomena, are complementary. They are mutually exclusive, but together they provide a complete description of the phenomena" \cite{bohr28}.}

Quantum mechanics also uniquely elevates the role of observation. Unlike previous theories, it does not explain how measurements produce definite outcomes. Without the collapse postulate, macroscopic superpositions arise. This is what later became known as the ``measurement problem,“ pictorially illustrated in terms of the Schrödinger's cat: a particle’s decay coupled to a cat could leave the cat in a superposition of alive and dead states, which is never observed.\footnote{Here are Schr\"odinger’s words: ``“One can even set up quite ridiculous cases. A cat is penned up in a steel chamber, along with the following device (which must be secured against direct interference by the cat): in a Geiger counter there is a tiny bit of radioactive substance, so small, that perhaps in the course of the hour one of the atoms decays, but also, with equal probability, perhaps none; if it happens, the counter tube discharges and through a relay releases a hammer which shatters a small flask of hydrocyanic acid. If one has left this entire system to itself for an hour, one would say that the cat still lives if meanwhile no atom has decayed. The first atomic decay would have poisoned it. The $\Psi$-function of the entire system would express this by having in it the living and the dead cat (pardon the expression) mixed or smeared out in equal parts” \cite{Schr}.}

The standard approach we saw in the first section solves this problem postulating that observation collapses the wave function into a definite state.
To underline the difficulty,  not to say the absurdity, of the collapse rule being related to our observations, Penrose gave the following argument during a debate at the 
British Institute of Art and Ideas\footnote{See https://iai.tv/video/quantum-and-the-unknowable-universe at min 21.}: Suppose that there is a planet far away from us, with no life and thus no observer--but that has an atmosphere which, according to quantum mechanics, is likely to be in a superposition of all kinds of states, a bit like Schrödinger's cat  but with many more terms in the sum than simply dead or alive. That is because the state of the atmosphere may depend on some microscopic events occurring in the particles composing it. 
Now, suppose that one sends a rocket to that planet; it takes a picture and that picture is sent back to earth; then someone looks at it; if one thinks that observation collapses the quantum state, then it is at that very moment that the superposition of atmospheres {\it on the distant planet} collapses into a single atmosphere!
 This is similar to the idea that looking at the cat modifies the state of the cat.
Penrose says (correctly) that it makes no sense at all; yet this follows logically from what students are told all over the world when they are learning quantum mechanics if the collapse rule is presented (as it often is) as being related to human observations.

 One way to avoid this problem is antirealism: the theory predicts observations but offers no understanding of an underlying reality. Von Neumann treated collapse epistemically: measurement updates our knowledge without changing the system physically.
Alternatively, some, like Wigner, suggested that consciousness plays an active role in the collapse, implying the world depends on our subjective experience. However, there is the risk of collapsing into solipsism if one does not want to accept the idea that looking here at a picture affects the state of the atmosphere on a distant planet. Physicists such as d’Espagnat, Zeilinger, and Mermin have sometimes entertained such views.\footnote{D’Espagnat: ``The doctrine that the world is made up of objects whose existence is independent of human consciousness turns out to be in conflict with quantum mechanics” \cite{SciAm}. Zeilinger: ``the distinction between reality and information cannot be made” \cite[p. 743]{Zeil}. Mermin: ``the moon is demonstrably not there when nobody looks” \cite[p. 397]{Merm}. To this, Stove replied: “[Mammals] depend for their existence on many things; but somebody’s looking at them is not among those things” \cite[pp. 99–100]{Stove}.} 

If one treats standard quantum mechanics as complete, the collapse postulate seems to force these extreme positions. However, one can also reject standard quantum mechanics as incomplete and seek alternatives that preserve realism, which is the case of the de Broglie-Bohm theory, where particles have definite positions and outcomes are determinate.\footnote{There exist other ``realist" approaches: the Ghirardi-Rimini-Weber (GRW) theory, introducing stochastic, spontaneous collapses as a law of nature \cite{grw}; the Everettian Many-Worlds, where all possible outcomes exist in different ``branches” \cite{Everett}.
All these approaches maintain an external world independent of observers but they suffer from serious difficulties, see \cite{Bri}.} 

\subsection{Scientific Realism}\label{sec2.1}
Scientific realism holds that our best scientific theories provide a reliable representation of a mind-independent world. It is a specialized application of general realism: while realism asserts that an external world exists, scientific realism claims that the unobservable entities posited by science—such as quarks, black holes, forces, and fields—also exist, roughly as described by theory.

Consider Newton’s theory of universal gravitation. Newton posited a force attracting bodies toward each other based on their mass, regardless of distance. By definition, this force is unobservable directly; we see only its effects on the motion of bodies, unlike contact forces. Newton introduced gravity because it accurately reproduced celestial motion, assuming it had physical reality while hoping for a deeper explanation in the future.\footnote{ Newton himself expressed caution: ``I have not as yet been able to discover the reason for these properties of gravity from phenomena, and I do not feign hypotheses” \cite[p. 943]{IN1}.}
Later developments in physics introduced other unobservable entities, such as electromagnetic waves, which propagate in a vacuum without a medium, and the curved geometry of space-time in general relativity, which is shaped by matter and energy.

Scientific realists argue that these entities exist because the corresponding mathematical structures provide the best explanation for empirical data. The alternative—that phenomena behave as if these entities exist but they do not—would be conspiratorial or miraculous.
In short, scientific realism maintains that the unobservable entities posited by science are real, and that their existence is justified by the explanatory power of the theories that describe them.

Some have proposed antirealism, restricting theories to observable quantities. This stance was considered by several pre-quantum physicists, such as Mach, but became especially influential after quantum mechanics. Yet this approach faces serious problems. 
First, observation is not independent of theory: as Einstein emphasized, relying solely on observations is naive: observation itself is a physical process governed by theory.\footnote{ He said to Heisenberg: ``But on principle, it is quite wrong to try founding a theory on observable magnitudes alone. In reality the very opposite happens. It is the theory which decides what we can observe. You must appreciate that observation is a very complicated process [\dots]. Only theory, that is, knowledge of natural laws, enables us to deduce the underlying phenomena from our sense impressions” \cite[pp. 63--64]{Heis3}.}

Second, the explanatory power of unobservable entities—such as gravitational forces, electromagnetic waves, or curved space-time—cannot be dismissed; denying their existence would require assuming a miraculous coordination of appearances. Third, the predictive success of theories invoking unobservables suggests that these entities correspond, at least approximately, to real features of the world. Historical continuity reinforces this point: scientific concepts often evolve rather than vanish entirely, indicating that prior theoretical entities were not mere fictions. Finally, in quantum mechanics, treating the wave function collapse as purely epistemic presupposes that the theory is incomplete, leaving the physical processes underlying measurement unexplained. Altogether, these considerations show that antirealism fails to account for the success, coherence, and continuity of scientific practice, supporting the realist view that unobservable entities are part of a mind-independent reality.

\subsection{Scientific Realism about Quantum Theory}\label{sec2.3}
We have seen that quantum theories compatible with scientific realism, like the de Broglie-Bohm theory, exist. But they contain 
a central feature: a highly mysterious mathematical object in the formalism, namely the wave function, which, as noted, does not vibrate in ordinary three-dimensional space. From a scientific realist perspective, how should we interpret it?

One might appeal to the same line of reasoning used for other unobservable entities: assuming the existence of the wave function provides the best explanation for the empirical data, much like assuming the existence of electromagnetic fields does. In other words, the experimental success of a theory—whether celestial mechanics, electromagnetism, atomic theory, relativity, or quantum mechanics—serves as evidence for the reality of the unobservable entities it posits. After all, we have never directly seen living dinosaurs, black holes, or the Sun’s interior, yet we infer their existence and properties from indirect observations. By the same logic, it seems natural to treat the wave function as real.
In quantum theory, this may lead to what is called ``wave function realism": Albert \cite{albert} and Ney \cite{ney} have independently argued that, by taking $\mathbb{R}^{N}$ (where $N$ is the number of degrees of freedom of the Universe) as the ``real’’ physical space, one can think of the wave function as a  physical field on that space.\footnote{For a review of alternative realist options about the wave function, see \cite{chen}, and for a critique of wave function realism,  see \cite{alloribook} and references therein.}

Historically, thinking of the world as high-dimensional and of matter as a field in that space was dismissed as obviously false, as Schrödinger, de Broglie, Lorentz, and Einstein all remarked. 
But, if we do not want to treat the wave function in a purely epistemic fashion, what should scientific realism about quantum mechanics commit us to?
The de Broglie–Bohm theory has an obvious answer: the world is made of particles, and the wave function describes their evolution.

\subsection{ Classical Reduction vs Quantum Mechanics } 
In Newtonian mechanics, fundamental entities are three-dimensional point particles, and macroscopic bodies are composed of them: gases consist of loosely interacting particles, while in liquids and solids interactions grow stronger. Macroscopic properties, such as the transparency of water, are explained in terms of microscopic dynamics. At each scale, one can identify pseudo-fundamental objects—atoms, molecules, etc.—that, while not truly fundamental, behave as if they were. This constitutes what are sometimes called the “special sciences”: they describe phenomena at a given scale as if the objects involved were fundamental.

Classically, however, the special sciences had to be reduced to physics, the most fundamental science. For example, thermodynamic laws reduce to mechanics via statistical mechanics, assuming gases are just a collection of particles. In this sense, classical reduction assumes that fundamental and higher-level entities inhabit the same space, with the former building up the latter, and microscopic dynamics determining macroscopic properties. Following Einstein’s terminology, statistical mechanics is a ``constructive” theory because it explains phenomena in terms of underlying entities and their dynamics, whereas thermodynamics is a ``principle” theory, accounting for phenomena through general constraints (e.g., energy conservation).\footnote{ 
About constructive theories, he wrote: ``They attempt to build up a picture of the more complex phenomena out of the materials of a relatively simple formal scheme from which they start out. Thus the kinetic theory of gases seeks to reduce mechanical, thermal, and diffusional processes to movements of molecules - i.e., to build them up out of the hypothesis of molecular motion. When we say that we have succeeded in understanding a group of natural processes, we invariably mean that a constructive theory has been found which covers the processes in question” (\cite{Einst1919}, republished in \cite{ideasop}, p. 227). 
About principle theories, he continues stating that ``the elements which form their basis and starting-point are not hypothetically constructed but empirically discovered ones, general characteristics of natural processes, principles which give rise to mathematically formulated criteria which the separate processes or the theoretical representations of them have to satisfy. Thus the science of thermodynamics seeks by analytical means to deduce necessary conditions, which separate events have to satisfy, from the universally experienced fact that perpetual motion is impossible.”}

As we said in the first section, ordinary quantum mechanics really deals only with predictions of results of laboratory experiments. It is neither a principle nor a constructive theory; in particular, it does not explain how measuring devices, made in principle of material particles, operate according to its own laws (hence the temptation to appeal to a non material consciousness to ``collapse" the wave function).
A constructive quantum theory  requires three-dimensional fundamental entities.
This is why one is led to “complete” quantum mechanics with a spatiotemporal ontology—most naturally the particle ontology of the de Broglie–Bohm theory.

The de Broglie-Bohm theory, thanks to its particle ontology, preserves reductionism. 
In that framework, the wave function is best seen as  a dynamical object determining particle motion, analogous in some respects to classical potentials or Hamiltonians.\footnote{For a view along these lines, see \cite{GZ} and \cite{alloriwf}.} 

\section{Conclusions}\label{sec5}
In this paper we have argued that it is only if one accepts the standard textbook view of quantum mechanics—with its dual laws of evolution for the wave function, the centrality of observations, and the mysterious collapse postulate—that the theory appears to undermine long-standing philosophical commitments. In that framework, one might conclude that realism must be abandoned, that reductionism is untenable, that determinism has been refuted, or even that quantum mechanics provides novel resources for solving the mind–body problem.

However, once we regard ordinary quantum mechanics as an incomplete description, and consider its natural completion in the de Broglie–Bohm theory, none of these radical claims holds. The de Broglie-Bohm completion restores a clear and coherent physical picture: it is a theory about matter in motion, governed by precise dynamical laws, much like classical mechanics or electromagnetism. On this view, quantum mechanics does not force us into metaphysical extravagance or radical revision of our most basic philosophical commitments.

It is true that the de Broglie–Bohm theory does not solve the mind–body problem, but this is not a peculiar shortcoming of the theory. No physical theory could achieve that, if one accepts the force of the so-called hard problem of consciousness. What the de Broglie-Bohm theory does provide, however, is a scientifically and philosophically well-grounded account of the physical world, free from the unnecessary mysteries that plague the orthodox formulation.

The main novelty of the de Broglie–Bohm framework is its explicit nonlocality, a feature already implicit in standard quantum mechanics but made transparent in de Broglie-Bohm terms (see \cite{Ma}). We have not addressed this issue in detail, as it does not affect the specific philosophical questions considered here. What matters for our purposes is that once we take the de Broglie-Bohm theory as the completed theory, quantum mechanics no longer appears as an enemy of realism, reductionism, or determinism. Rather, it takes its place alongside other physical theories as part of our best scientific account of the natural world.

\end{document}